\RequirePackage{lineno}

\documentclass[aps,prc,twocolumn,groupedaddress,showpacs,amsmath,amssymb,floatfix,superscriptaddress]{revtex4}
\usepackage{multirow}
\usepackage{graphicx}
\usepackage{times}
\usepackage{color}

\newcommand{\PerTonDay}{(ton~day)$^{-1}$}

\begin{document}

\title{Measurement of the double-$\beta$ decay half-life of $^{136}$Xe with the KamLAND-Zen experiment}

\newcommand{\tohoku}{\affiliation{Research Center for Neutrino
    Science, Tohoku University, Sendai 980-8578, Japan}}
\newcommand{\osaka}{\affiliation{Graduate School of 
    Science, Osaka University, Toyonaka, Osaka 560-0043, Japan}}
\newcommand{\lbl}{\affiliation{Physics Department, University of
    California, Berkeley, and \\ Lawrence Berkeley National Laboratory, 
Berkeley, California 94720, USA}}
\newcommand{\colostate}{\affiliation{Department of Physics, Colorado
    State University, Fort Collins, Colorado 80523, USA}}
\newcommand{\ut}{\affiliation{Department of Physics and
    Astronomy, University of Tennessee, Knoxville, Tennessee 37996, USA}}
\newcommand{\tunl}{\affiliation{Triangle Universities Nuclear
    Laboratory, Durham, North Carolina 27708, USA and \\
Physics Departments at Duke University, North Carolina Central University,
and the University of North Carolina at Chapel Hill}}
\newcommand{\ipmu}{\affiliation{Institute for the Physics and Mathematics of the 
    Universe, University of Tokyo, Kashiwa 277-8568, Japan}}
\newcommand{\nikhef}{\affiliation{Nikhef and the University of Amsterdam, Science Park, Amsterdam, the Netherlands}}
\newcommand{\washington}{\affiliation{Center for Experimental Nuclear Physics and Astrophysics, University of Washington, Seattle, Washington 98195, USA}}

%
%
\author{A.~Gando}\tohoku
\author{Y.~Gando}\tohoku
\author{H.~Hanakago}\tohoku
\author{H.~Ikeda}\tohoku
\author{K.~Inoue}\tohoku\ipmu
\author{R.~Kato}\tohoku
\author{M.~Koga}\tohoku\ipmu
\author{S.~Matsuda}\tohoku
\author{T.~Mitsui}\tohoku
\author{T.~Nakada}\tohoku
\author{K.~Nakamura}\tohoku\ipmu
\author{A.~Obata}\tohoku
\author{A.~Oki}\tohoku
\author{Y.~Ono}\tohoku
\author{I.~Shimizu}\tohoku
\author{J.~Shirai}\tohoku
\author{A.~Suzuki}\tohoku
\author{Y.~Takemoto}\tohoku
\author{K.~Tamae}\tohoku
\author{K.~Ueshima}\tohoku
\author{H.~Watanabe}\tohoku
\author{B.D.~Xu}\tohoku
\author{S.~Yamada}\tohoku
\author{H.~Yoshida}\tohoku

\author{A.~Kozlov}\ipmu

\author{S.~Yoshida}\osaka

\author{T.I.~Banks}\lbl
\author{J.A.~Detwiler}\lbl
\author{S.J.~Freedman}\ipmu\lbl
\author{B.K.~Fujikawa}\ipmu\lbl
\author{K.~Han}\lbl
\author{T.~O'Donnell}\lbl

\author{B.E.~Berger}\colostate

\author{Y.~Efremenko}\ipmu\ut

\author{H.J.~Karwowski}\tunl
\author{D.M.~Markoff}\tunl
\author{W.~Tornow}\tunl

\author{S.~Enomoto}\ipmu\washington

\author{M.P.~Decowski}\ipmu\nikhef

\collaboration{KamLAND-Zen Collaboration}\noaffiliation

\date{\today}

\begin{abstract}

We present results from the KamLAND-Zen double-beta decay experiment based on an exposure of 77.6~days with 129~kg of $^{136}$Xe. The measured two-neutrino double-beta decay half-life of $^{136}$Xe is $T_{1/2}^{2\nu} = 2.38 \pm 0.02({\rm stat}) \pm 0.14({\rm syst}) \times 10^{21}$~yr, consistent with a recent measurement by \mbox{EXO-200}. We also obtain a lower limit for the neutrinoless double-beta decay half-life, $T_{1/2}^{0\nu} > 5.7 \times 10^{24}$~yr at 90\% confidence level~(C.L.), which corresponds to almost a five-fold improvement over previous limits.
\end{abstract}


\maketitle

\section{Introduction}
Majorana neutrinos are a natural feature of many high-energy physics theoretical models. However, the only viable experimental probe of this property at present is neutrinoless double-beta ($0\nu\beta\beta$) decay~\cite{Gomez2012}. Observation of this lepton-number violating nuclear process would definitively establish the Majorana nature of the neutrino, and would be a profound discovery~\cite{Schechter1982}. In addition, since the rate of this process increases with the square of the effective neutrino mass $\left<m_{\beta\beta}\right> \equiv \left| \Sigma_{i} U_{ei}^{2}m_{\nu_{i}} \right|$, its measurement would provide information on the absolute neutrino mass scale. Searches for $0\nu\beta\beta$ decay have been invigorated by recent measurements of neutrino mass splittings by oscillation experiments, which require at least one neutrino mass above $\sim$50~meV~\cite{Nakamura2010}. This scale is within the reach of present-day efforts.

Determining $\left<m_{\beta\beta}\right>$ from a $0\nu\beta\beta$ decay half-life requires knowledge of the decay's phase-space factor ($G^{0\nu}$) and nuclear matrix element ($M^{0\nu}$). $G^{0\nu}$ can be calculated exactly, but to date all estimations of $M^{0\nu}$ must rely on model-based approximations with difficult-to-quantify uncertainties. The two-neutrino double-beta ($2\nu\beta\beta$) decay half-life, if known, can be used to constrain some relevant model parameters, reducing some sources of uncertainty ~\cite{Rodin2003, Simkovic2008}. The first direct measurement of the $2\nu\beta\beta$ decay half-life of $^{136}$Xe,  recently reported by \mbox{EXO-200} \cite{Ackerman2011}, was significantly below previously published lower limits~\cite{Bernabei2002, Gavriljuk2006}. This article on the first results from the KamLAND-Zen (KamLAND Zero-Neutrino Double-Beta Decay) experiment reports a new measurement of the $^{136}$Xe $2\nu\beta\beta$ decay half-life, as well as improved limits on the $0\nu\beta\beta$ mode. The data presented were collected between October 12, 2011, and January 2, 2012.

\section{Detector and calibration}
\begin{figure}[t]
\begin{center}
\includegraphics[angle=0,width=1.0\columnwidth]{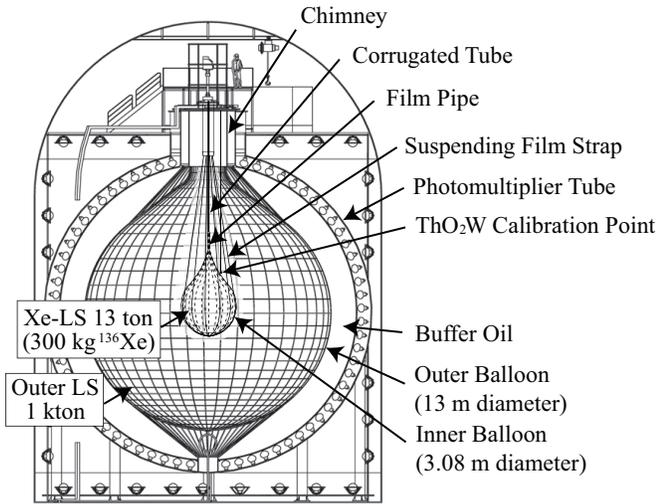}
\vspace{-0.8cm}
\end{center}
\caption[]{Schematic diagram of the KamLAND-Zen detector.}
\label{figure:detector}
\end{figure}

KamLAND-Zen (Fig.~\ref{figure:detector}) is a modification of the existing KamLAND detector carried out in the summer of 2011. The $\beta\beta$ source/detector is \mbox{13~tons} of Xe-loaded liquid scintillator~(Xe-LS) contained in a 3.08-m-diameter spherical inner balloon (IB). The IB is constructed from 25-$\mu$m-thick transparent nylon film and is suspended at the center of the KamLAND detector \cite{Abe2010} by 12 film straps of the same material.  The IB is surrounded by 1 kton  of liquid scintillator (LS) contained in a 13-m-diameter spherical outer balloon (OB) made of 135-$\mu$m-thick nylon/EVOH (ethylene vinyl alcohol copolymer) composite film. The outer LS is 0.10\% less dense than the Xe-LS and acts as an active shield for external $\gamma$'s and as a detector for internal radiation from the Xe-LS or IB. The Xe-LS consists of 82\% decane and 18\% pseudocumene (1,2,4-trimethylbenzene) by volume, 2.7 g/liter of the fluor PPO (2,5-diphenyloxazole), and $(2.52 \pm 0.07)$~\% by weight of enriched xenon gas, as measured by gas chromatography. The isotopic abundances in the enriched xenon were measured by residual gas analyzer to be $(90.93 \pm 0.05)\%$ \mbox{$^{136}$Xe} and $(8.89 \pm 0.01)\%$ \mbox{$^{134}$Xe}; other xenon isotopes are negligible. The light yield of the Xe-LS is 3\% lower than that of the LS. Buffer oil (BO) between the OB and an 18-m-diameter spherical stainless-steel containment tank (SST) shields the LS from external radiation. Scintillation light is recorded by 1,325 17-inch and 554 20-inch photomultiplier tubes (PMTs) mounted on the SST, providing 34\% solid-angle coverage. The SST is surrounded by a \mbox{3.2~kton} water-Cherenkov outer detector (OD). Details of the KamLAND detector are given in Ref.~\cite{Abe2010}.

The data acquisition system (DAQ) is triggered when 70 or more 17-inch PMTs are hit (primary trigger), which corresponds to a threshold of $\sim$0.4~MeV. The signals on all hit PMTs are digitized for $\sim$200~ns for offline analysis. After each primary trigger the threshold is lowered to $\sim$0.25~MeV for 1~ms to study sequential decays. The scintillation light from the two coincident $e^{-}$ produced by $^{136}$Xe $\beta\beta$ decay cannot be separated, so only their summed energy is observed. For hypothetical $0\nu\beta\beta$ decays, the sum is always 2.458~MeV (\mbox{\it{Q}~\rm{value}} of the $^{136}$Xe $\beta\beta$ decay) ~\cite{Redshaw2007}, while for the $2\nu\beta\beta$ decays the sum has a continuous spectrum up to the \mbox{\it{Q}~\rm{value}}.
Event energy (visible energy) is estimated from the number of observed photoelectrons (p.e.) after correcting for PMT gain variation and solid angle, shadowing, and transparency of detector materials. The corrections depend on the event vertex. The vertex reconstruction is based on the maximum likelihood fit to the pulse shape of each PMT hit timing after correcting for photon time of flight. The pulse shape is almost determined by scintillation decay time and dark hit contribution, and it differs between 17-inch and 20-inch PMTs due to different transit-time spreads.
The vertex resolution is estimated from radial distributions of radioactive contaminants (see Fig.~\ref{figure:attenuation}) to be $\sigma\sim$$15~{\rm cm}/\sqrt{E({\rm MeV})}$. The energy response is calibrated with $\gamma$'s from a $^{208}$Tl (\mbox{${\rm ThO_{2}W}$}) source, $^{214}$Bi ($\beta + \gamma$'s) from $^{222}$Rn (\mbox{$\tau = 5.5$~day}) introduced during detector modification, and 2.225~MeV $\gamma$'s from spallation neutrons captured on protons. 

\begin{figure}[t]
\begin{center}
\includegraphics[width=1.05\columnwidth]{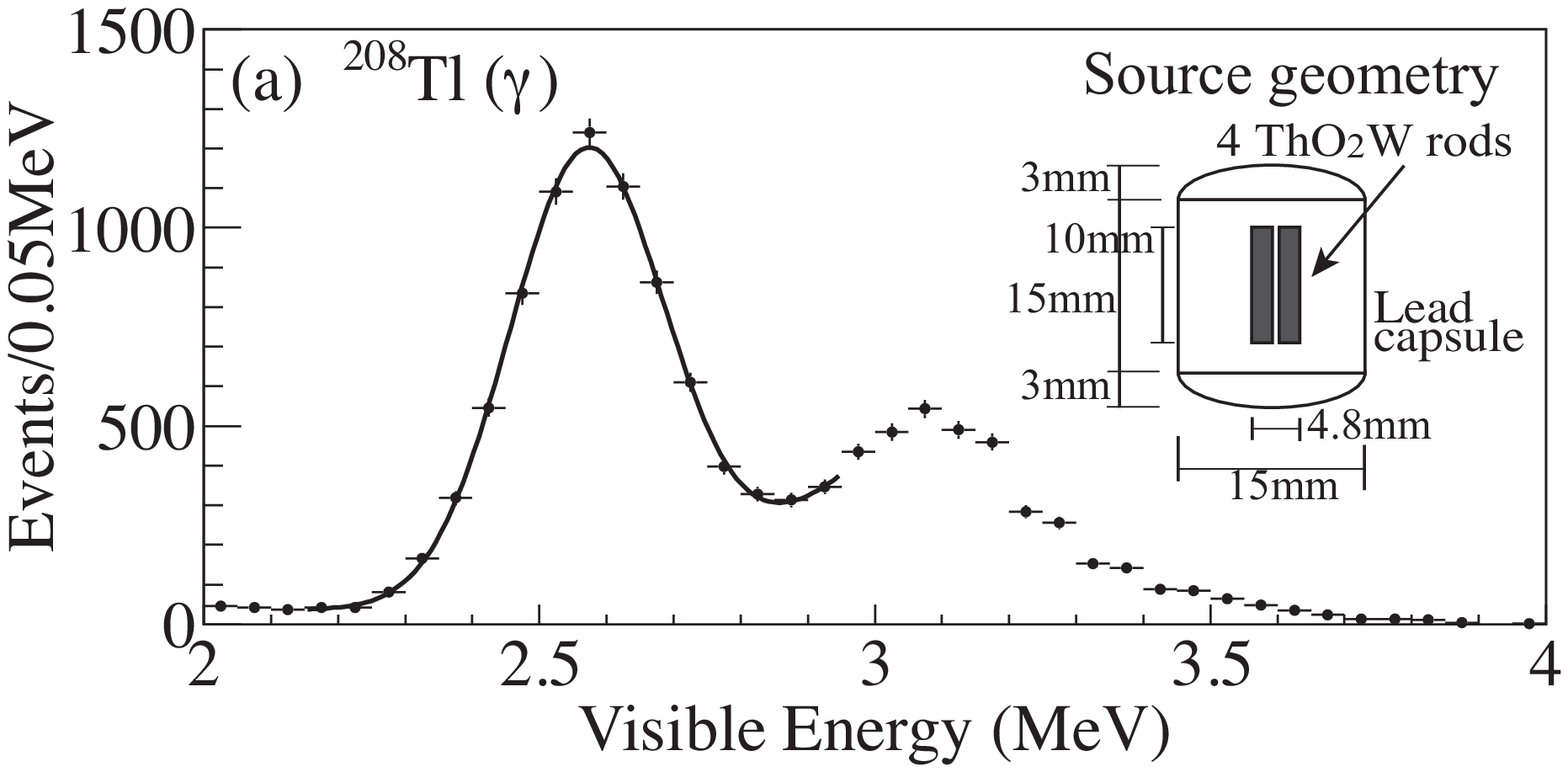}
\vspace{-1.0cm}
\end{center}
\begin{center}
\includegraphics[width=1.05\columnwidth]{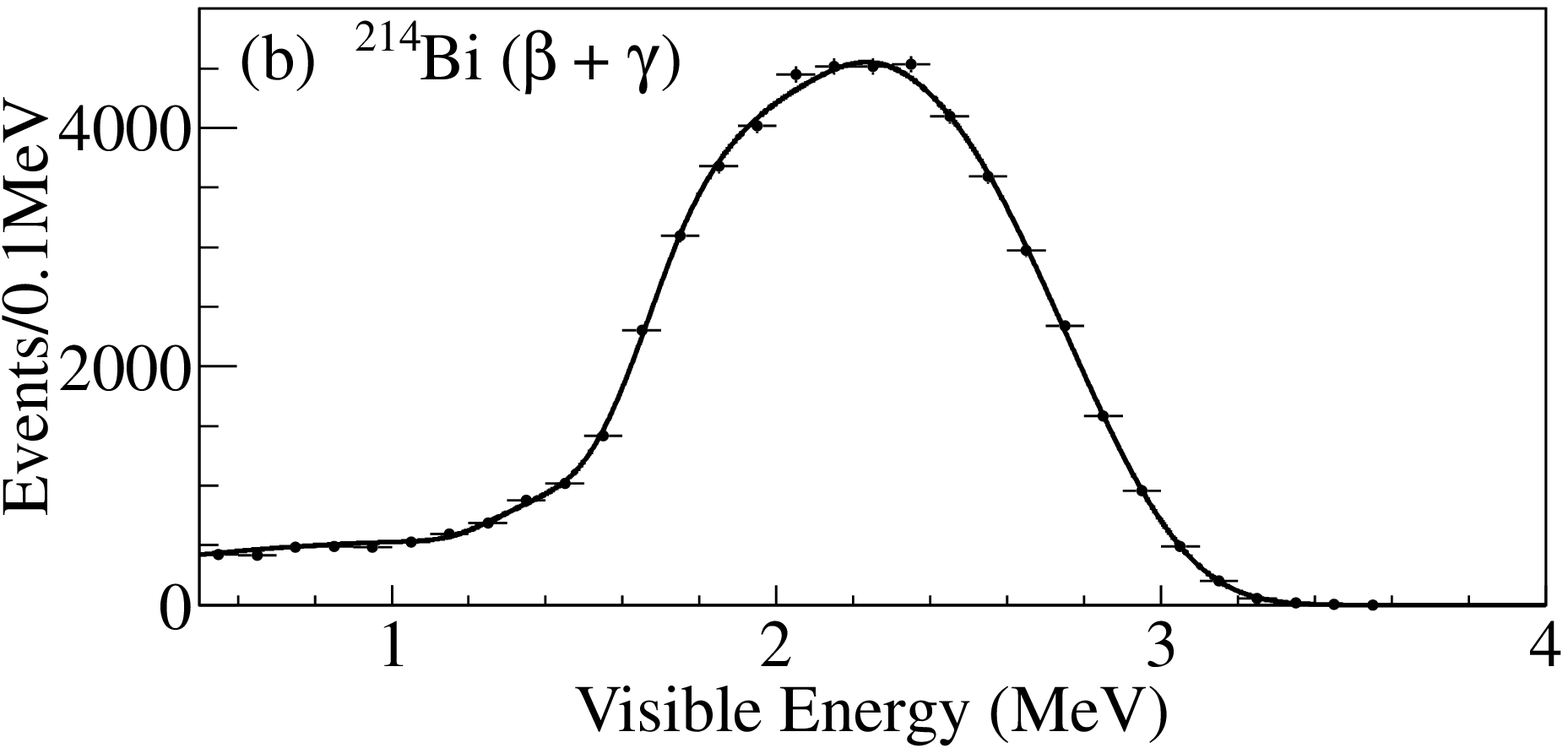}
\vspace{-0.7cm}
\end{center}
\caption[]{Visible energy distributions of (a) $\gamma$'s from the \mbox{$^{208}$Tl} calibration source and (b) $^{214}$Bi ($\beta + \gamma$'s) decays in the Xe-LS. The lines indicate the best fits to the analytical spectral models with the resolution and energy scale parameters floating. The fit to (a) has a $\chi^2$/d.o.f. = 5.0/8 and (b) has a $\chi^2$/d.o.f. = 27.0/29.}
\label{figure:calibration}
\end{figure}

Figure~\ref{figure:calibration}(a) shows the energy spectrum obtained when the \mbox{${\rm ThO_{2}W}$} source, contained in a $\sim$5-mm-thick lead capsule, was deployed close to the outer surface of the IB. The most intense peak is due to the primary $\gamma$ of $^{208}$Tl (2.614~MeV). The less intense peak near $\sim$3.1~MeV is from multiple-$\gamma$ cascades of $^{208}$Tl. According to Monte Carlo (MC) studies, the degradation of the primary $\gamma$ inside the source is negligible, and the distribution around the primary peak can be described by a Gaussian distribution and a third-order polynomial. The mean and width of the Gaussian distribution are relatively insensitive to the polynomial parameters. The resultant energy resolution at 2.614~MeV is $\sigma$ = $(6.6 \pm 0.3)\%/\sqrt{E({\rm MeV})}$. The parameters of a detector energy nonlinear response model describing effects from scintillator quenching and Cherenkov light production are constrained  to reproduce the 2.614~MeV $^{208}$Tl peak position and the spectral shape of $^{214}$Bi events [Fig.~\ref{figure:calibration}(b)]. From the neutron-capture $\gamma$ data, systematic variation of the energy reconstruction over the Xe-LS volume is less than 1.0\%, and the detector energy response is stable to within 1.0\% during the data set. 

\section{Candidate event selection}
\begin{figure}[]
\begin{center}
\includegraphics[width=1.1\columnwidth]{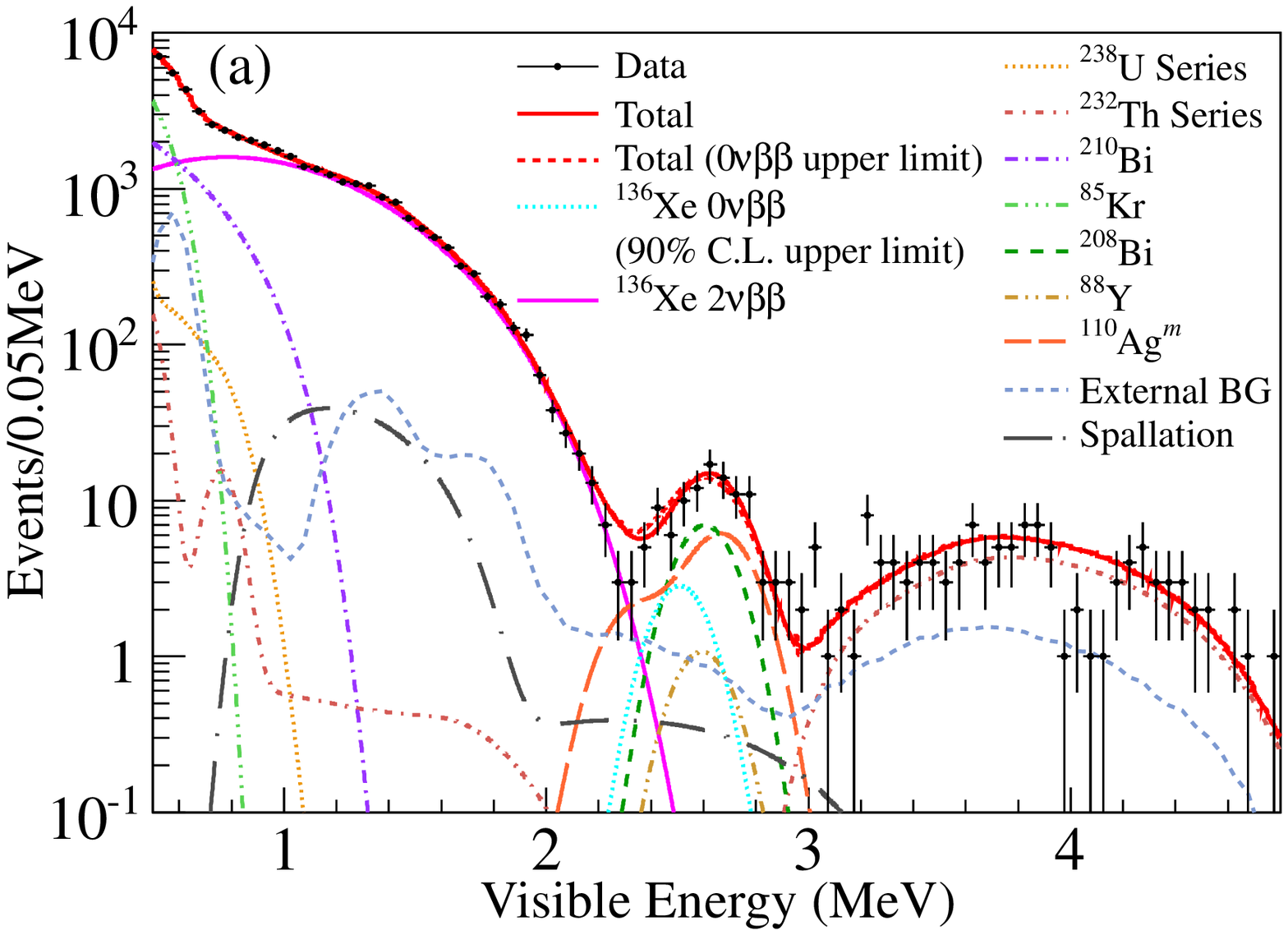}
\vspace{-0.7cm}
\end{center}
\begin{center}
\includegraphics[width=1.1\columnwidth]{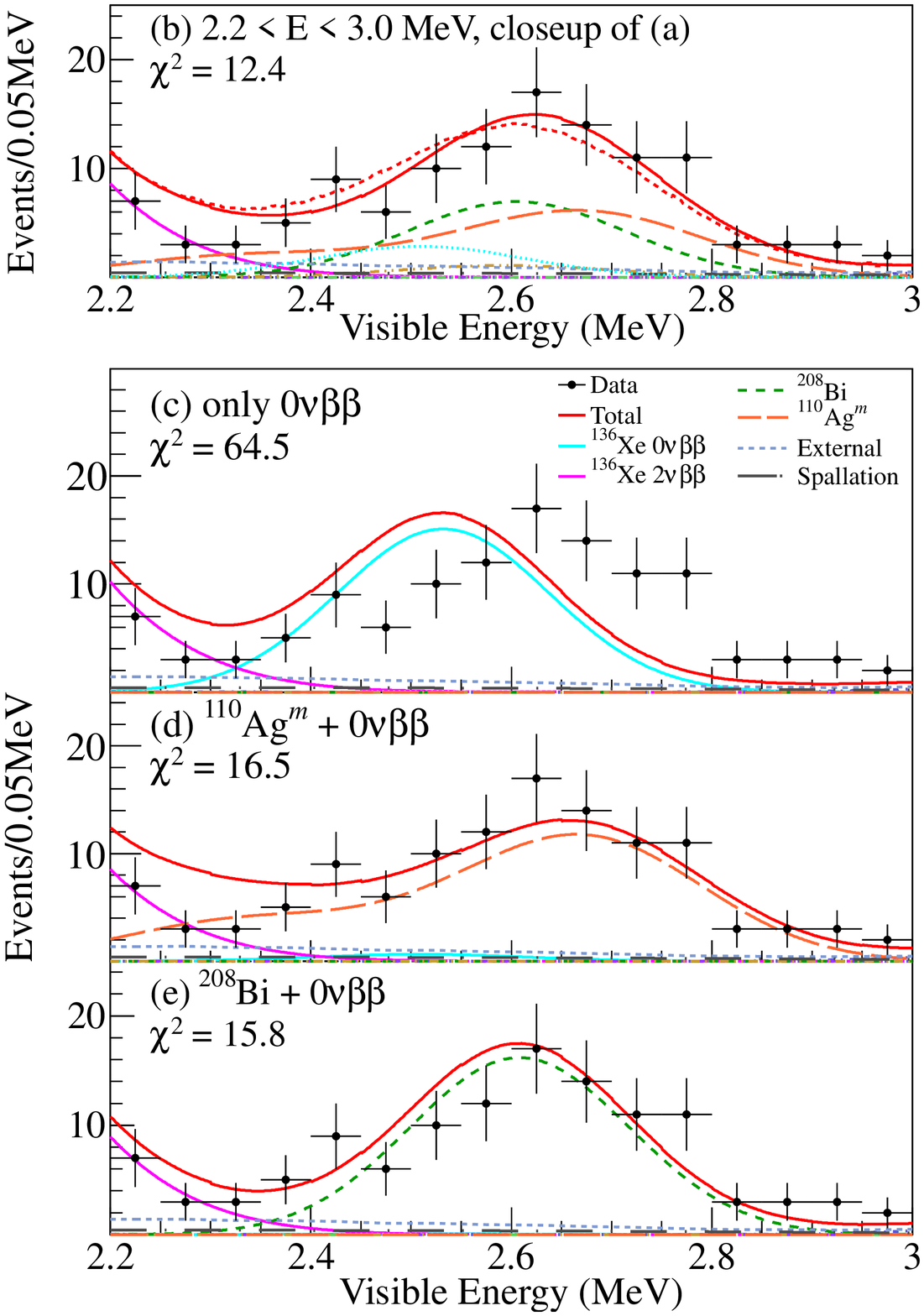}
\vspace{-0.7cm}
\end{center}
\caption[]{ (Color) (a) Energy spectrum of selected $\beta\beta$ decay candidates together with the best-fit backgrounds and $2\nu\beta\beta$ decays, and the 90\% C.L. upper limit for $0\nu\beta\beta$ decays; the fit range is $0.5~{\rm MeV} < E < 4.8~{\rm MeV}$. The coincident $\beta/\gamma$ events from $^{208}$Tl give the main contribution  from 3 to 5 MeV. (b) Closeup of (a) for $2.2~{\rm MeV} < E < 3.0~{\rm MeV}$. (c)-(e) Closeups for $2.2~{\rm MeV} < E < 3.0~{\rm MeV}$ with different background assumptions around the $0\nu\beta\beta$ peak: (c) only 0$\nu\beta\beta$, (d) $^{\rm 110}$Ag$^m$ + 0$\nu\beta\beta$, (e) $^{208}$Bi+ 0$\nu\beta\beta$, for comparison. $\chi^2$ are from the data in this energy range.}
\label{figure:energy}
\end{figure}

Candidate $\beta\beta$ decay events are selected by performing the following series of cuts: (i) The reconstructed vertex must be within 1.2~m of the detector center, defining the fiducial volume (FV). (ii) Muon (events with more than 10,000 p.e. or more than 5 OD hits) and events occurring within 2~ms after muons are eliminated. (iii) A coincidence cut eliminates sequential events that occur within 3~ms of each other; this removes $(99.97 \pm 0.01)\%$ of $^{214}$Bi-$^{214}$Po ($\beta+\gamma$, then $\alpha$ decay, $\tau$=237\,$\mu$s) decays. This cut is augmented with a secondary cut, aimed at identifying sequential $^{212}$Bi-$^{212}$Po ($\tau$=0.4\,$\mu$s), which exploits detailed PMT wave-form data to identify coincidences within a single $\sim$200-ns-long DAQ event window. The $^{212}$Bi-$^{212}$Po rejection efficiency is $(89 \pm 2)\%$. The visible energy of $\alpha$ decay is quenched due to the high ionization density, so the cuts are applied only if the delayed $\alpha$'s are between 0.35 and 1.5 MeV. The dead time introduced by the coincidence cuts is less than $\sim$0.1\%. (iv) A background mainly from reactor $\overline{\nu}_{e}$'s producing a delayed coincidence of positrons and neutron-capture $\gamma$'s (2.225 MeV) is rejected by requiring event time separations greater than 1~ms, and the delayed energy is larger than 1.5 MeV. (v) Finally, candidates must pass a vertex-time-charge (VTQ) test designed to filter out noise events. The test compares the observed PMT charge and hit-time distributions to those expected based on the reconstructed vertex~\cite{Abe2011}. The VTQ cut is tuned using KamLAND calibration data and reduces the selection efficiency by less than 0.1\%. The total livetime after all cuts is 77.6 days. The energy spectrum of $\beta\beta$ decay candidates is shown in Fig.~\ref{figure:energy}.

\section{Background estimation}
\begin{figure}[t]
\begin{center}
\includegraphics[width=1.0\columnwidth]{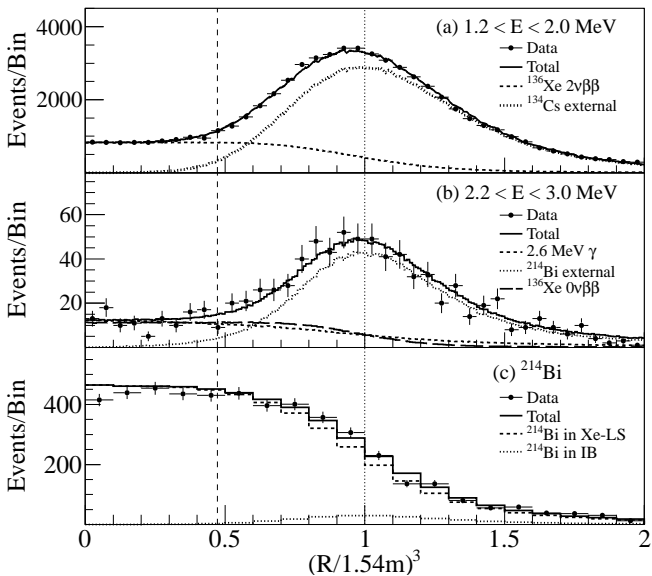}
\vspace{-0.7cm}
\end{center}
\caption[]{$R^{3}$ vertex distribution of candidate events for (a) $1.2~{\rm MeV} < E < 2.0~{\rm MeV}$ and (b) $2.2~{\rm MeV} < E < 3.0~{\rm MeV}$. The curves show the best-fit model components: (a) $2\nu\beta\beta$ (dashed) and $^{134}$Cs (dotted). (b) 2.6~MeV $\gamma$'s (dashed) and $^{214}$Bi (dotted); the long-dashed curve is for $0\nu\beta\beta$ instead of $\gamma$'s. (c) $^{214}$Bi events from Xe-LS (dashed) and IB (dotted). The vertical lines show the fiducial radius of 1.2 m (dashed) and the IB radius (dotted).}
\label{figure:attenuation}
\end{figure}

Backgrounds to the $\beta\beta$ decay study fall into three categories: those external to the Xe-LS, mainly from the IB material; those from residual radioactive impurities in the Xe-LS; and spallation backgrounds~(Table \ref{tb:bg}). From a spectral analysis of events whose reconstruction is close to the IB boundary, we find that the activity in the energy region $1.2~{\rm MeV} < E < 2.0~{\rm MeV}$ ($2\nu\beta\beta$ window) is dominated by $^{134}$Cs ($\beta + \gamma$); in the region $2.2~{\rm MeV} < E < 3.0~{\rm MeV}$ ($0\nu\beta\beta$ window), the spectrum is consistent with $^{214}$Bi ($\beta + \gamma$) decays. The observed surface activity ratio of $^{134}$Cs to $^{137}$Cs (0.662 MeV $\gamma$) is $\sim$0.8 which is consistent with contamination by fallout from the Fukushima-I reactor accident in March 2011. The IB production facility is located just 100 km away from the Fukushima-I reactor. The FV cut is performed to mitigate the background from the IB material; the remaining IB background inside the FV is estimated by fitting Monte-Carlo-generated event radial distributions to the data. Figure~\ref{figure:attenuation} shows the event density as a function of the cubed radius from the IB center for the two energy ranges, along with fits to the MC distributions. In the $2\nu\beta\beta$ window we fit for a $2\nu\beta\beta$ source uniformly distributed in the Xe-LS and a $^{134}$Cs source uniformly distributed on the IB.  In the $0\nu\beta\beta$ window we show the best fits for a $^{214}$Bi source uniformly distributed on the IB and either a $0\nu\beta\beta$-like source or a 2.6~MeV $\gamma$ source uniformly distributed in the Xe-LS. The radial distribution offers no discrimination between these event types.    

\begin{table}[htdp]
\caption[]{Summary of background in the fiducial volume. Note that $^{134}$Xe $\beta\beta$ contribution is expected to be negligible because of much smaller phase space factor and isotopic abundance relative to $^{136}$Xe.
The backgrounds from nuclear reactions such as ($\alpha$, $\gamma$) and  ($\alpha$, $\alpha\gamma$) are expected to be negligible due to their small cross sections. The rate of ($n$, $\gamma$) is stringently constrained from the data.\\
}
\label{tb:bg}

\begin{center}
\begin{tabular}{c c c}
\hline
\hline
\multicolumn{2}{c}{Isotope}&Event rate ~\PerTonDay\\
\hline
\\
\multicolumn{3}{c}{External (radioactivity in IB)}\\
\hline
$^{238}$U series&$^{222}$Rn-$^{210}$Pb&$(3.2 \pm 0.3) \times 10^{-2}$\\
$^{232}$Th series&$^{228}$Th-$^{208}$Pb&$(3.5 \pm 0.3) \times 10^{-2}$\\
\multicolumn{2}{c}{$^{40}$K}&$0.21 \pm 0.16$\\
\multicolumn{2}{c}{$^{134}$Cs}&$0.50 \pm 0.07$\\
\multicolumn{2}{c}{$^{137}$Cs}&$0.35 \pm 0.06$\\
\\
\multicolumn{3}{c}{Residual radioactivity in Xe-LS}\\
\hline
\multirow{2}{*}{$^{238}$U series}&$^{238}$U-$^{222}$Rn ($^{234}$Pa)&$<$1.5\\
&$^{222}$Rn-$^{210}$Pb&4.9 $\pm$ 0.2 \\
\multirow{2}{*}{$^{232}$Th series}&$^{232}$Th-$^{228}$Th ($^{228}$Ac)&$<$0.7\\
&$^{228}$Th-$^{208}$Pb&0.58 $\pm$ 0.06 \\
\multicolumn{2}{c}{$^{85}$Kr}&$196 \pm 8$\\
\multicolumn{2}{c}{$^{210}$Bi}&$103 \pm 3$\\
\multicolumn{2}{c}{$^{40}$K}&$<$9.6\\
\multicolumn{2}{c}{$^{134}$Cs}&$<$0.4\\
\multicolumn{2}{c}{$^{90}$Y}&$<$0.8\\
\multicolumn{2}{c}{$^{137}$Cs}&$<$1.1\\
\\
\multicolumn{3}{c}{Spallation product from $^{12}$C $^a$} \\
\hline
\multicolumn{2}{c}{$^{10}$C}&$(2.11 \pm 0.44) \times 10^{-2}$\\
\multicolumn{2}{c}{$^{11}$C}&$1.11 \pm 0.28$\\
\\
\multicolumn{3}{c}{Spallation products from xenon with lifetime $<$ 100\,s}\\
\hline
\multicolumn{2}{c}{$1.2~{\rm MeV} < E < 2.0~{\rm MeV}$}&$<$ 0.3\\
\multicolumn{2}{c}{$2.2~{\rm MeV} < E < 3.0~{\rm MeV}$}&$<$ 0.02\\
\\
\multicolumn{3}{c}{Around $^{136}$Xe 0$\nu\beta\beta$ (ENSDF search)}\\
\hline
\multicolumn{2}{c}{$^{60}$Co, $^{88}$Y, $^{\rm 110}$Ag$^m$ and $^{208}$Bi}&$0.22 \pm 0.04$\\
\\
\multicolumn{3}{c}{Potential background from fallout}\\
\hline
\multicolumn{2}{c}{$^{134}$Cs, $^{137}$Cs, $^{\rm 110}$Ag$^m$}&listed above\\
\multicolumn{2}{c}{$^{\rm 129}$Te$^m$, $^{95}$Nb, $^{90}$Y, $^{89}$Sr}&negligible\\
\hline
\hline
\end{tabular}
\end{center}
\begin{flushleft}
$^a$ Other spallation products are negligible \cite{Abe2010}.\\
\end{flushleft}
\end{table}

Assuming secular equilibrium, the residual $^{238}$U and $^{232}$Th concentrations internal to the Xe-LS are estimated to be $(3.5 \pm 0.6) \times 10^{-16}$~g/g and $(2.2 \pm 0.3) \times 10^{-15}$~g/g, respectively, based on sequential decays of $^{214}$Bi-$^{214}$Po and $^{212}$Bi-$^{212}$Po.  Since equilibrium may be broken by introduction of contaminants during detector modification, the Bi-Po studies are only used to estimate internal background from the $^{222}$Rn-$^{210}$Pb subchain of the $^{238}$U series, and from the $^{228}$Th-$^{208}$Pb subchain of the $^{232}$Th series; other decays in both series are treated as unconstrained backgrounds. We note the well-known 2.614~MeV $\gamma$ from $^{208}$Tl ($\beta^{-}$ decay, \mbox{$Q=5.00$~MeV}) in the $^{232}$Th series is not a serious background for the $0\nu\beta\beta$ decay search because of detection of the coincident $\beta/\gamma$ in the surrounding active LS~\cite{Raghavan1994}. 

Spallation neutrons are tagged by coincidence of neutron-capture $\gamma$'s with preceding muons. We expect capture on protons (2.225~MeV), $^{12}$C (4.946~MeV), $^{136}$Xe (4.026~MeV), and $^{134}$Xe (6.364~MeV), with fractions 0.994, 0.006, $9.5 \times 10^{-4}$, and $9.4 \times 10^{-5}$, respectively. We find no $^{136}$Xe or $^{134}$Xe neutron-capture candidates in the data set. $^{137}$Xe ($\beta^{-}$, \mbox{$\tau=5.5$~min}, \mbox{$Q=4.17$~MeV}) from neutron capture on $^{136}$Xe is a potential $0\nu\beta\beta$ background, but the expected production rate is negligible, $\sim$$2.9 \times 10^{-3}$~\PerTonDay, where ton is a unit of Xe-LS mass. Production rates of light nuclei by spallation of carbon are calculated from spallation yields previously measured in KamLAND ~\cite{Abe2010}. We observe a $(13 \pm 6)$\% increase in the spallation neutron flux in the Xe-LS relative to the outer LS, from which we assess a 19\% systematic uncertainty on the calculated spallation yields. Events from decays of $^{11}$C ($\beta^{+}$, \mbox{$\tau=29.4$~min}, \mbox{$Q=1.98$~MeV}) and $^{10}$C ($\beta^{+}$, \mbox{$\tau = 27.8$~s}, \mbox{$Q = 3.65$~MeV}) dominate the contributions from spallation backgrounds. We expect rates of $1.11 \pm 0.28$~\PerTonDay\, and $(2.11 \pm 0.44) \times 10^{-2}$~\PerTonDay\, from $^{11}$C and $^{10}$C, respectively. The $^{11}$C/$^{10}$C background can be reduced by a triple-coincidence tag of a muon, a neutron, and the $^{11}$C/$^{10}$C decay. This is not pursued in the current analysis. We found no past experimental data for muon spallation of xenon. With the present data, we find the event rates in the energy ranges $1.2~{\rm MeV} < E < 2.0~{\rm MeV}$ and $2.2~{\rm MeV} < E < 3.0~{\rm MeV}$ from isotopes with lifetimes of less than 100\,s associated with muons depositing more than $\sim$3\,GeV in the detector (so-called showering muons) are less than 0.3~\PerTonDay and 0.02~\PerTonDay\, at 90\% C.L., respectively.

Care has to be taken for backgrounds that produce a peak close to the $0\nu\beta\beta$ decay energy (2.458 MeV), particularly ones which may have been introduced during detector modification or may be induced by muon spallation.  We searched all isotopes in the ENSDF database~\cite{ENSDF2006} for sources with a peak structure between 2.4 and 2.8 visible-MeV.  For all tabulated decay chains we calculate the visible energy spectrum, accounting for the time structure of the chain and pileup in the DAQ event window, and apply the nonlinear detector energy response model to all individual decay secondaries of each branch. Considering only nuclei with decay ancestor lifetimes longer than 30~days, we identify $^{\rm 110}$Ag$^m$ ($\beta^{-}$ decay, \mbox{$\tau=360$~day}, \mbox{$Q = 3.01$~MeV}), $^{88}$Y (electron capture~(EC) decay, \mbox{$\tau=154$~day}, \mbox{$Q = 3.62$~MeV}), $^{208}$Bi (EC decay, \mbox{$\tau=5.31 \times 10^{5}$~yr}, \mbox{$Q = 2.88$~MeV}), and $^{60}$Co ($\beta^{-}$ decay, \mbox{$\tau = 7.61$~yr}, \mbox{$Q = 2.82$~MeV}) as potential background sources.  Observation of $^{134}$Cs/$^{137}$Cs on the IB raises the plausibility of contamination of detector materials by Fukushima fallout, which includes $^{\rm 110}$Ag$^m$. One assay of soil samples taken near the IB production facility revealed evidence of $^{\rm 110}$Ag$^m$. Although $^{88}$Y, $^{208}$Bi, and $^{60}$Co are not detected near Fukushima or in our soil samples, we consider them to be possible backgrounds. Except for $^{208}$Bi, these long-lived background candidates can be also produced from xenon spallation by cosmic-rays when materials were aboveground, but the rate estimations are difficult. Broadening the search to include shorter-lived nuclei ($100~{\rm s} < \tau < 30~{\rm day}$) possibly supported by muon spallation in the detector, we found that the production of candidate parents with mass numbers below $^{136}$Xe is stringently constrained by comparing production cross sections in Ref.~\cite{Napolitani2007}. 

\section{Systematic uncertainty}
Nominally, the 1.2-m-radius FV corresponds to $0.438 \pm 0.005$ of the total Xe-LS volume ($16.51 \pm 0.17~{\rm m}^{3}$), or 129~kg of $^{136}$Xe. The fiducial volume fraction may also be estimated from the fraction of $^{214}$Bi events whose reconstruction is within 1.2~m of the IB center compared to the total number in the entire Xe-LS volume after subtraction of the IB surface contribution. The result is $0.423 \pm 0.007({\rm stat}) \pm 0.004({\rm syst})$ [Fig.~\ref{figure:attenuation}(c)], including the uncertainty on the IB surface contribution. The slight event loss around the center may indicate an outward radial vertex bias, so the difference in these estimates is taken as a measure of the  systematic error on the vertex-defined FV. Combining the errors, we obtain a 5.2\% systematic error on the fiducial volume.  The total systematic uncertainty on the $\beta\beta$ decay half-life measurement is 5.9\%, coming from the quadrature sum of the fiducial volume (5.2\%), enrichment of $^{136}$Xe (0.05\%), Xe concentration (2.8\%), detector energy scale (0.3\%), Xe-LS edge effect (0.06\%), and detection efficiency (0.2\%). 

\section{Result}
The $^{136}$Xe $2\nu\beta\beta$ and $0\nu\beta\beta$ decay rates are estimated from a likelihood fit to the binned energy spectrum of the selected candidates between 0.5 and 4.8 MeV. The $^{136}$Xe $2\nu\beta\beta$ decay spectrum shape from Ref.~\cite{Haxton1984} is used; 80\% of the $2\nu\beta\beta$ spectrum falls within our energy window. The contributions from major backgrounds in the Xe-LS, such as $^{85}$Kr, $^{40}$K, nonequilibrium $^{210}$Bi, and the $^{238}$U-$^{222}$Rn and $^{232}$Th-$^{228}$Th decay chains, are free parameters and are left unconstrained in the fit. The contributions from the $^{222}$Rn-$^{210}$Pb and $^{228}$Th-$^{208}$Pb chains, $^{11}$C, and $^{10}$C are allowed to vary but are constrained by their independent measurements. Residual IB-surface backgrounds in the FV are constrained by the radial distribution study. The parameters of the detector energy response model are floated but are constrained to reproduce the $^{208}$Tl source and $^{222}$Rn-induced $^{214}$Bi data. Potential backgrounds from fallout nuclei with half-lives longer than 30~days found in $ex~situ$ measurements of soil or ocean samples around Fukushima, namely $^{137}$Cs, $^{134}$Cs, $^{\rm 110}$Ag$^m$, $^{\rm129}$Te$^m$, $^{95}$Nb, $^{90}$Y (from $^{90}$Sr), and $^{89}$Sr, as well as potential $0\nu\beta\beta$ backgrounds found in the ENSDF search ($^{88}$Y, $^{208}$Bi, and $^{60}$Co) are included as unconstrained free parameters. The relative contributions of $0\nu\beta\beta$ window backgrounds are additionally constrained by the time variation of the event rate in the energy range $2.2~{\rm MeV} < E < 3.0~{\rm MeV}$.

Figure~\ref{figure:energy}(a) shows the resulting best-fit spectral decomposition. The $\chi^{2}$/d.o.f. comparing the binned data and the best-fit expectation is 99.7/87. $2\nu\beta\beta$ decay is the dominant spectral feature in the low-energy region. The best-fit number of $^{136}$Xe $2\nu\beta\beta$ decays is $(3.55 \pm 0.03) \times 10^{4}$, corresponding to an event rate of $80.9 \pm 0.7$~\PerTonDay. We found no systematic variations due to the choice of the data period and volume within the 1.2-m-radius FV. The dominant backgrounds at low-energy are from $^{85}$Kr and $^{210}$Bi, with best-fit rates of $196 \pm 8$~\PerTonDay and $103 \pm 3$~\PerTonDay, respectively. The fit yields the following 90\% C.L. upper limits on other background rates (per ton$\cdot$day) in the Xe-LS: $^{40}$K $<$9.6, $^{234}$Pa$<$1.5, $^{134}$Cs$<$0.4, $^{228}$Ac $<$0.7, $^{90}$Y $<$ 0.8, and $^{137}$Cs$<$1.1; other fallout isotopes are negligible. 

 Around the $0\nu\beta\beta$ energy, a strong peak appears, but the peak is centered significantly above the \mbox{\it{Q} \rm{value}} of the decay [Fig.~\ref{figure:energy}(c)]; the hypothesis that the peak can be described by $0\nu\beta\beta$ decay alone is rejected by a $\chi^{2}$ test at more than $5\sigma$ C.L., including the systematic uncertainties on the energy scale model. The best-fit combined background rate around the $0\nu\beta\beta$ energy allowing for contributions from $^{\rm 110}$Ag$^m$, $^{88}$Y, $^{208}$Bi, and $^{60}$Co is $0.22 \pm 0.04$~\PerTonDay. Figures~\ref{figure:energy}(d) and \ref{figure:energy}(e) show the distribution if only $^{208}$Bi or $^{\rm 110}$Ag$^m$, respectively, contribute as background to the $0\nu\beta\beta$ peak. We conclude that the data in the $0\nu\beta\beta$ region is contaminated by the candidate backgrounds (but mainly due to $^{208}$Bi or $^{\rm 110}$Ag$^m$), and the $0\nu\beta\beta$ limit is extracted by floating those contributions [Fig.~\ref{figure:energy}(b)]. The 90\% C.L upper limit on the number of $^{136}$Xe $0\nu\beta\beta$ decays is $<$ 15~events, an event rate of $<$ 0.034~\PerTonDay. 

The measured $2\nu\beta\beta$ decay half-life of $^{136}$Xe is $T_{1/2}^{2\nu} = 2.38 \pm 0.02({\rm stat}) \pm 0.14({\rm syst}) \times 10^{21}$~yr. This is consistent with the result obtained by \mbox{EXO-200}, $T_{1/2}^{2\nu} = 2.11 \pm 0.04({\rm stat}) \pm 0.21({\rm syst}) \times 10^{21}$~yr~\cite{Ackerman2011}. For $0\nu\beta\beta$ decay, the data give a lower limit of $T_{1/2}^{0\nu} > 5.7 \times 10^{24}$~yr (90\% C.L.), which corresponds to almost a fivefold improvement over previous limits~\cite{Bernabei2002}. From the limit on the $^{136}$Xe $0\nu\beta\beta$ decay half-life we obtain a 90\% C.L. upper limit of $\left<m_{\beta\beta}\right> < (0.3-0.6)~{\rm eV}$ using recent QRPA (CCM SRC)~\cite{Simkovic2009} and shell model~\cite{Menendez2009} nuclear matrix elements calculated prior to the \mbox{EXO-200} measurement.

\section{Conclusion}
In summary, KamLAND-Zen provides an improved measurement of the $^{136}$Xe $2\nu\beta\beta$ decay half-life. The result is consistent with that of \mbox{EXO-200} and supports the conclusion that the directly measured half-life is significantly less than the lower limits reported by earlier experiments. Our analysis includes a search for $0\nu\beta\beta$ decay of $^{136}$Xe and yields an improved lower limit on its half-life. Removal of contaminants in the Xe-LS is an important task to improve the $0\nu\beta\beta$ decay search sensitivity. In the future, systematic uncertainties will also be reduced by performing source calibrations in the Xe-LS. 

\section*{ACKNOWLEDGMENTS}
The KamLAND-Zen experiment is supported by the Grant-in-Aid for Specially Promoted Research under Grant No.~21000001 of the Japanese Ministry of Education, Culture, Sports, Science and Technology; the World Premier International Research Center Initiative (WPI Initiative), MEXT, Japan; and under the US Department of Energy (DOE) Grant No. DE-AC02-05CH11231, as well as other DOE grants to individual institutions. The Kamioka Mining and Smelting Company has provided service for activities in the mine.

\bibliography{DoubleBeta}

\end{document}